\documentclass[aps,pra,groupedaddress,superscriptaddress,twocolumn,10pt,longbibliography]{revtex4-2}
\pdfoutput=1 
\usepackage[utf8]{inputenc}
\usepackage{hyperref}
\usepackage{graphicx}
\usepackage{orcidlink}
\usepackage{booktabs}
\usepackage{tabularx}
\usepackage{cancel}

\usepackage{comment}

\newcommand{\cmtout}[1]{}

\usepackage{siunitx} 
\newcommand{\omatt}[1]{}

\usepackage{soul} 
\usepackage[normalem]{ulem}
\usepackage{amssymb}
\usepackage{physics}
\usepackage{hyperref}
\DeclareSIUnit\bohr{\text{\ensuremath{a_\textup{0}}}}

\makeatletter
\def\maketitle{
\@author@finish
\title@column\titleblock@produce
\suppressfloats[t]}
\makeatother
\begin{document}

\title{Observation of $\Pi$-symmetry ultralong-range Rydberg molecules}

\author{Matthew T. Eiles
\orcidlink{0000-0002-0569-7551}}
\thanks{These authors contributed equally to this work.}
\affiliation{Department of Physics and Astronomy, Purdue University, West Lafayette, IN 47907, USA}
\affiliation{Max Planck Institute for the Physics of Complex Systems,  Nöthnitzer Str. 38, 01187 Dresden, Germany}

\author{Aleksandr Zaitsev}
\thanks{These authors contributed equally to this work.}
\affiliation{Institut f\"{u}r Quantenmaterie and Center for Integrated Quantum Science \\ and Technology IQ$^{ST}$, Universit\"{a}t Ulm, D-89069 Ulm, Germany}

\author{Dominik Dorer \orcidlink{0000-0002-6988-7506}}
\affiliation{Institut f\"{u}r Quantenmaterie and Center for Integrated Quantum Science \\ and Technology IQ$^{ST}$, Universit\"{a}t Ulm, D-89069 Ulm, Germany}
\author{Shinsuke Haze
\orcidlink{0000-0003-1696-3947}
}
\affiliation{Institut f\"{u}r Quantenmaterie and Center for Integrated Quantum Science \\ and Technology IQ$^{ST}$, Universit\"{a}t Ulm, D-89069 Ulm, Germany}
\affiliation{Center for Quantum Information and Quantum Biology, The University of Osaka, 1-2 Machikaneyama, Toyonaka, Osaka 560-0043, Japan}
\author{Markus Deiß\orcidlink{0009-0003-8025-910X}}
\affiliation{Institut f\"{u}r Quantenmaterie and Center for Integrated Quantum Science \\ and Technology IQ$^{ST}$, Universit\"{a}t Ulm, D-89069 Ulm, Germany}

\author{S. Efe G\"urleyen\orcidlink{0000-0003-3363-3202}}
\affiliation{Department of Physics and Astronomy, Purdue University, West Lafayette, IN 47907, USA}
\author{Chris H. Greene
\orcidlink{0000-0002-2096-6385}}
\affiliation{Department of Physics and Astronomy, Purdue University, West Lafayette, IN 47907, USA}
\affiliation{Purdue Quantum Science and Engineering Institute, Purdue University, West Lafayette, IN 47907, USA}
\author{Johannes Hecker Denschlag\orcidlink{0000-0003-1984-4994}}
\affiliation{Institut f\"{u}r Quantenmaterie and Center for Integrated Quantum Science \\ and Technology IQ$^{ST}$, Universit\"{a}t Ulm, D-89069 Ulm, Germany}

\date{\today}

\begin{abstract}
We observe weakly-bound $\Pi$-symmetry electronic states in the spectroscopy of $^{87}$Rb$(nP_{3/2})$+$^{87}$Rb($5S_{1/2}$) ultralong-range Rydberg molecules.
We detect these molecules in Rydberg states having principal quantum number $13\le n \le 16$. 
Their $\Pi$-state character is unambiguously identified via their observed multiplet structure: the $2F+1$ magnetic sublevels of the ground-state rubidium atom separate, as in the Zeeman effect, because of the spin-spin coupling between the Rydberg and valence electrons.
We find a rapid decrease in the molecular binding energy $\propto (n-\mu_{P_{3/2}})^{-11}$, where $\mu_{P_{3/2}}$ is the quantum defect, indicating that the low-$n$ regime of Rydberg states is ideally suited for studies of $\Pi$-symmetry molecules. 
 Our observations are in good agreement with Green's function-based calculations for $14\le n\le 16$, with poorer agreement for $n=13$ hinting at the beginning of a breakdown of the Fermi pseudopotential approach at low $n$.
\end{abstract}

\maketitle

Due to its rotational symmetry, a diatomic molecule  generally exhibits a conserved angular momentum about its internuclear axis. 
For tightly-bound molecules
where the electrons are strongly coupled to the axis, 
the projection ($m_\ell$) of the  electronic orbital angular momentum $\ell$
 is typically a good quantum number.  In molecular spectroscopy, $\Sigma$-states $(\Lambda\equiv |m_\ell| = 0)$ and $\Pi$-states $(\Lambda = 1)$ are routinely observed with binding strengths of similar magnitude \cite{Herzberg}.
For loosely-bound molecules, however, other interactions, such as spin-orbit coupling, can mix different $\Lambda$-states. 

A prime example of such molecules are ultralong-range Rydberg molecules (ULRMs) formed from a Rydberg atom and a distant ground-state ``perturber" atom \cite{
greene2000,
Sadeghpour2018, eiles_trilobites_2019, fey_ultralong-range_2020, Yoshida2024}.
The Rydberg electron responsible for molecular binding potentially experiences 
not only spin-orbit coupling, but also coupling to the electronic and nuclear spins of the perturber \cite{khuskivadze2002adiabatic,anderson2014photoassociation,eilesHamiltonian2017,rivera2024approximate,niederprum2016rydberg,peper2020photodissociation,kleinbach2017photoassociation,Boettcher2016,deis_observation_2020}. 
Therefore, $\Lambda$ is generally not 
a good quantum number for ULRMs, apart from exceptions, e.g., when spin-orbit coupling vanishes \cite{greene2000,Althoen2023,exner_high_2025}.
As we show here, for a pure $\Pi$-state ULRM it is
the weakness of the electron-atom interaction that effectively prevents coupling to other $\Lambda$-states. As a further consequence of their weak interactions,
$\Pi$-state ULRMs exhibit extremely small binding energies, which explains why their observation has been elusive so far.

\begin{figure}[b]
\centering
\includegraphics[width=0.48\textwidth]{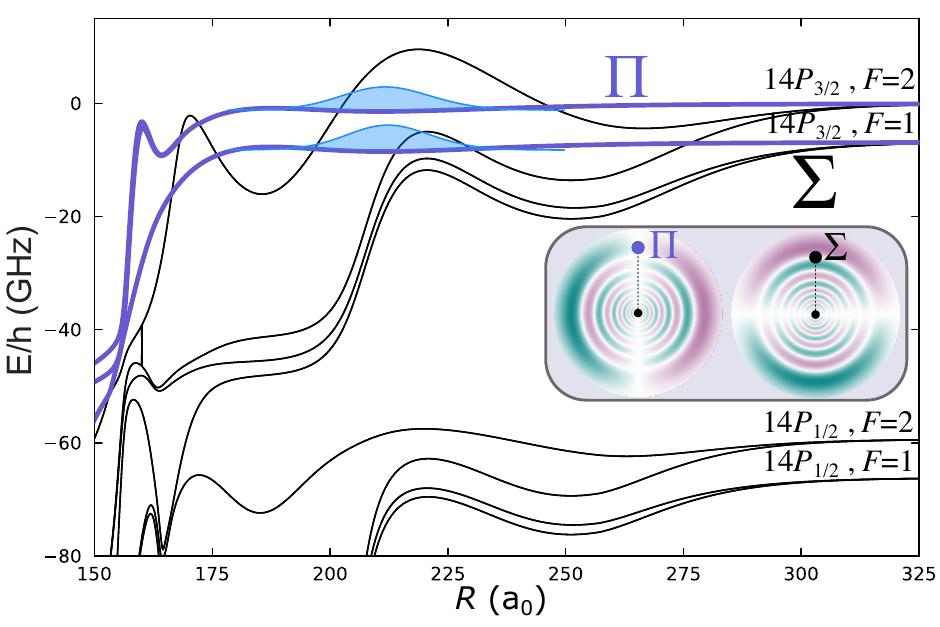}
\caption{
Potential energy curves (PECs)
for $^{87}$Rb $14P_j$ ultralong-range Rydberg molecules. 
Purple (black) curves correspond to 
$\Pi$ ($\Sigma$) states. 
The vibrational ground-state wave functions for the $\Pi$-states are shown in blue.  The inset shows the $nP$ electron orbital for $\Pi$ and $\Sigma$ molecular states (left and right, respectively). For a $\Pi$-state ($\Sigma$-state) molecule, the perturber (purple and black circles, respectively) sits in a region where only the gradient component perpendicular 
(parallel) to the internuclear axis is non-zero.
The PECs shown have $\Omega = 1/2$, but those for other $\Omega$ values look very similar.
}
\label{fig:intro}
\end{figure}

\begin{figure*}[t]
\centering
\includegraphics[width=\textwidth]{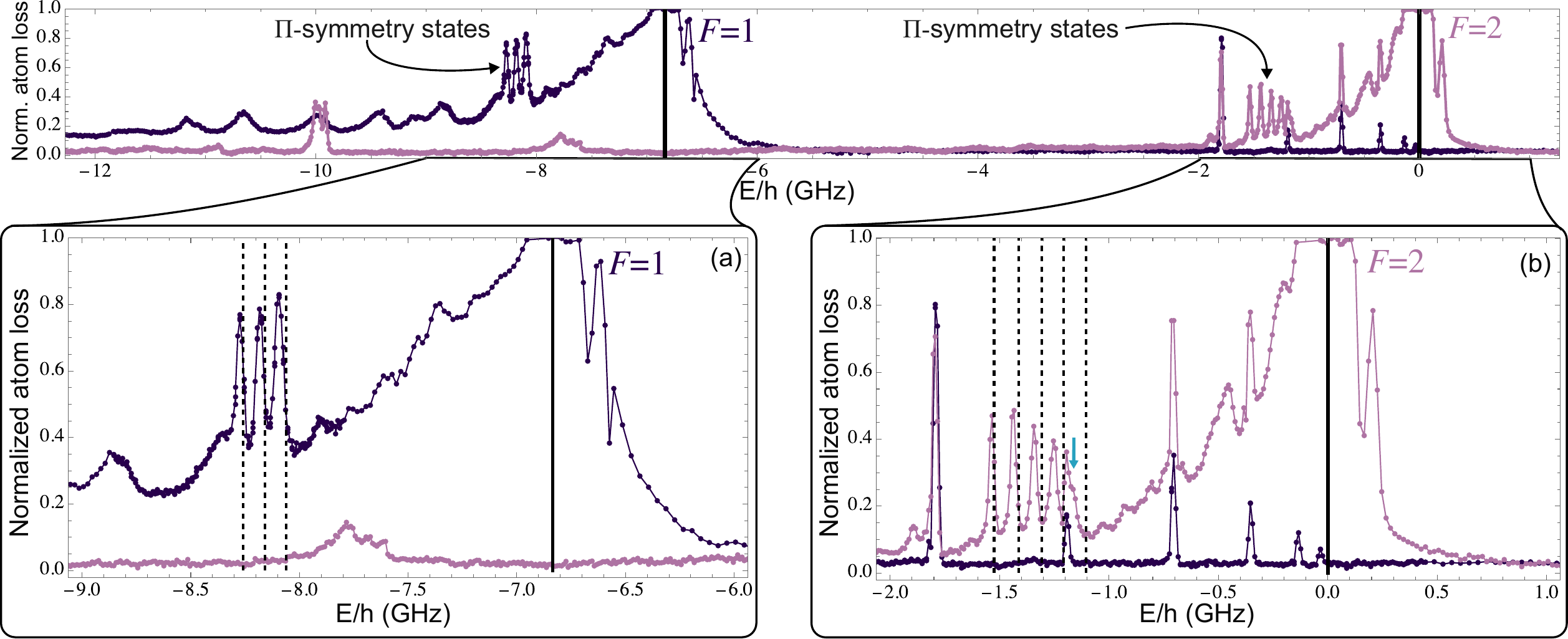}
\caption{ULRM spectra in the vicinity of the $14P_{3/2}$ Rydberg state for $^{87}$Rb atoms initially prepared in the $F=1$ (dark shade) and $F=2$ (light shade) state. Shown is the normalized atomic loss as a function of the {laser frequency} detuning 
{E/h}
of the UV spectroscopy laser from the $5S_{1/2}(F=2) \rightarrow 14P_{3/2}$ atomic transition. The solid vertical lines denote the $F=1$ and $F=2$ atomic thresholds. For better comparison, the spectra for atoms initially prepared in the $F=1$ ground state were shifted by twice the ground-state hyperfine splitting $(-6.834\, \mathrm{ GHz} \times \mathrm{h})$. The insets (a), (b) show zooms into the spectral regions around the spectra of the $\Pi$ molecular states. The positions, splittings, and number of states observed are in convincing agreement with our theoretical calculations (dashed vertical lines).
The fifth level of the $F=2$ series (marked by the blue arrow) is observed as a shoulder of the narrow peak at -1.2 GHz which, as it also appears in the $F=1$ spectrum, is assigned to an excited vibrational state in one of the $\Sigma$-state PECs (see also Fig. S1 in the Supplemental Material \cite{Supplemental}).
}
\label{fig:spectra}
\end{figure*}

In this Letter, we report the observation of pure $\Pi$-symmetry $^{87}$Rb$(nP_{3/2})+^{87}$Rb$(5S_{1/2})$ ultralong-range Rydberg molecules.
Their ground vibrational levels appear in characteristic multiplets exhibiting the same structure as the Zeeman effect of the perturber atom. 
We find that the binding energies and equidistant splittings within each multiplet both scale approximately as $ \nu^{-11}$, where $\nu = n - \mu_{P_{3/2}}$ is the effective principal quantum number with $\mu_{P_{3/2}}\approx 2.64$ the atomic quantum defect. 
This scaling is much more dramatic than the $\nu^{-6}$ scaling of $\Sigma$ molecular states observed in several atomic species \cite{Pfau2014,guttridge_individual_2025,desalvo_ultra-long-range_2015,sasmannshausen_experimental_2015,legrand2025revealing}, and reflects the comparatively much smaller binding energies of $\Pi$-state ULRMs. 
We will show in the theoretical analysis below that the weakness of these $\Pi$-state interactions originates in the geometry of the $P$-orbital, which has a node along the internuclear axis $\vec R$  [see inset of Fig.~\ref{fig:intro}].
We study Rydberg states with relatively low principal quantum numbers $13\le n\le 16$ where the $\Pi$-state multiplets are clearly resolved.

Figure \ref{fig:intro} shows a subset of the  molecular potential energy curves (PECs) for a $^{87}$Rb$(14P_{j=1/2,3/2})$ Rydberg atom and a $^{87}$Rb$(5S_{1/2},F=1,2)$ ground-state atom at large internuclear distance $R$.
Here, $j$ ($\vec{j} \equiv \vec{s}_e  + \vec{\ell}$) and $F$ are the total angular momenta
of the Rydberg electron and  of the ground-state atom, respectively, 
and $\vec{s}_e$ is the spin of the Rydberg electron.
The corresponding  
axis projections are $m_j, m_s, m_\ell$, and $M_F$.
The total angular momentum projection $\Omega = M_F + m_j$
is a strictly conserved quantity (when ignoring molecular rotation).
The black PECs correspond to states of predominantly $\Sigma$ symmetry, which possess comparatively deep minima at $R\sim 250\,a_0$ 
\footnote{To obtain smooth PECs for $R\sim 2n^2$ we have used electron-Rb scattering phase shifts obtained from a modified version of the model potential proposed in Ref.~\cite{chibisov2002energies}. Within the pseudopotential formalism, the long-range polarization tail of the Rb + e$^-$ interaction leads to unphysical divergences in the $p$-wave contribution when treated using zero-range pseudopotentials. We avoid these by truncating the polarization potential at $R\sim 100a_0$.  Further details of this {procedure} will be presented in a later publication as it is only relevant for the primarily $\Sigma$-symmetry PECs. The $\Pi$-symmetry PECs investigated here are independent of this truncation.}. 
Pure $\Pi$-states clearly exist for (pure) stretched states with $j = 3/2$ and $m_j = m_s + m_\ell = \pm 3/2$, because then $m_\ell = \pm 1$.  
The corresponding PECs are shown in purple. 
Although appearing nearly flat at this scale, these $\Pi$-state PECs possess shallow minima at $R\sim 210$ $a_0$, and thus we expect weakly-bound molecular states to be observable in the close vicinity of the $nP_{3/2}$ $F=1$ and $F=2$ thresholds. 

Therefore, we have carried out high precision spectroscopy in this region.
Fig. \ref{fig:spectra} shows the experimentally observed ULRM 
spectra in the vicinity of the $14P_{3/2},F=1 $ and $ 2$ pair state thresholds. 
The dark (light) spectrum was measured in a cloud of $^{87}$Rb atoms initially prepared in the $F=1$ ($F=2$) hyperfine level, respectively. 
Besides the two very broad and dominant 
atomic Rydberg resonance peaks, we observe
several narrow peaks which correspond to 
various  vibrational bound states supported by the PECs shown in Fig.~\ref{fig:intro}.
Among those scattered lines, we observe
in each spectrum a strikingly equidistant
multiplet about 1.5 GHz detuned from the 
respective atomic resonance (as marked in Fig. \ref{fig:spectra}).
For $F=2$ this multiplet has five peaks, while for $F=1$ it has three. 
 As we show below, each level of this multiplet can be unambiguously assigned to the vibrational ground state of a molecular PEC corresponding to an electronic $\Pi$-state. The splitting  arises from the breaking of the $(2F+1)$-fold degeneracy of the magnetic sublevels of the perturber by the coupling of {the Rydberg electron's spin $\vec s_e$ to the electronic spin $\vec s_p$ of the perturber atom.}
 The calculated level positions (dashed lines) relative to the atomic threshold agree with experiment to within 5\% relative error. 

In contrast to the multiplets, many of the other  observed narrow resonance lines appear simultaneously in both spectra. 
This is typical  for $\Sigma$-symmetry states where the interaction is so large that $F$ states can be mixed (see Fig.~\ref{fig:intro}). 
This mixing permits photoassociation of the same molecular state from either of the $F=1$ or $F=2$ atomic samples.

To produce the ULRMs, we prepared an ultracold cloud of $^{87}\mathrm{Rb}$ atoms at a temperature of approximately $1~\mu\mathrm{K}$ in a crossed optical dipole trap, operating at a wavelength of $1064~\mathrm{nm}$ and having a potential depth of about $20~ \mathrm{\mu K}$.  
With the trapping frequencies $\omega_{z,r} \approx 2 \pi \times(25,180)~\mathrm{Hz}$ and atom number $N \approx 3.6 \cdot 10^6$ we obtain a peak density $n_0 \approx 5.2 \times 10^{13}\, \mathrm{cm}^{-3}$. 
Atoms are selectively prepared in either the $(F = 1, M_F=-1)$ or $(F = 2, M_F=2)$ hyperfine level of the $5S_{1/2}$ ground state. 
ULRMs are formed by photoassociation using direct single-photon excitation with a narrow-linewidth laser (see also \cite{deis_observation_2020}) operating in the wavelength range of $302$--$306~\mathrm{nm}$. The laser beam is focused onto the atomic sample with a beam waist of approximately $100~\mu\mathrm{m}$ and has linear polarization.
The excitation duration and corresponding laser power are varied in the ranges of $200\text{--}800~\mathrm{ms}$ and $0.1\text{--}10~\mathrm{mW}$, respectively.
Resonant excitation of ULRMs is detected by measuring atom-number loss of the cloud via absorption imaging. By varying slightly our experimental scheme we can optimize our experiment 
in terms of spectral resolution or sensitivity of our spectroscopy (see End Matter).

In order to calculate the PECs, we use the Coulomb Green's function \cite{eiles2023} to compute the eigenenergies of the electronic Hamiltonian
\begin{equation}
\label{eq:hamiltonian}
    \hat H_e(\vec r;\vec R) = \hat H_\mathrm{Ryd}(\vec r) + \hat H_\mathrm{HF} + \hat V_e(\vec r;\vec R)
\end{equation}
for a given internuclear distance $\vec{R}$.
The Rydberg atom's Hamiltonian $\hat H_\mathrm{Ryd}$ possesses the eigenenergies $E_{n(\ell s_e)j}$ and eigenstates 
$|n(\ell s_e)jm_j\rangle$, while the hyperfine Hamiltonian of the ground-state atom $\hat H_\mathrm{HF}$ has eigenenergies $E_F$ and eigenstates $|(is_p)FM_F\rangle$,
where $i$ is the nuclear spin of the perturber atom.
The interaction between the electron and the ground-state Rb atom, $\hat V_e(\vec r,\vec R)$, is provided by the Fermi-Omont pseudopotential
\cite{Supplemental}
 parametrized by the singlet and triplet $s$-wave scattering lengths $a_{(^1S_0)}(K)$, $a_{(^3S_1)}(K)$, and $p$-wave scattering volumes $a_{(^1P_1)}^3(K)$, $a_{(^3P_J)}^3(K)$ \cite{anderson2014photoassociation,khuskivadze2002adiabatic,eilesHamiltonian2017}. 
 Here, $^{2S+1}L_J$ denotes the term symbol of the Rb $ + \, e^-$ complex and $K$ is the kinetic energy of the electron. 
 We use a model potential to describe the interaction between the Rb atom and a free electron \cite{khuskivadze2002adiabatic}. 
 With this we obtain the functions $a^{2L+1}_{(^{2S+1}L_J)}(K)$ consistent with the fitted values reported in Refs. \cite{engelPrecision2019,exner_high_2025,exnerObservation2026}.
 The $R$-dependent eigenvalues of $\hat H_e$ are the adiabatic PECs for each Rydberg level considered here (Fig.~\ref{fig:intro}). 
 Using these, we calculate the ULRM bound-state energies following the numerical method outlined in Refs.~\cite{durst_nonadiabatic_2025,guttridge_individual_2025}. 

\begin{figure}[t]
\centering
\includegraphics[width=0.47\textwidth]{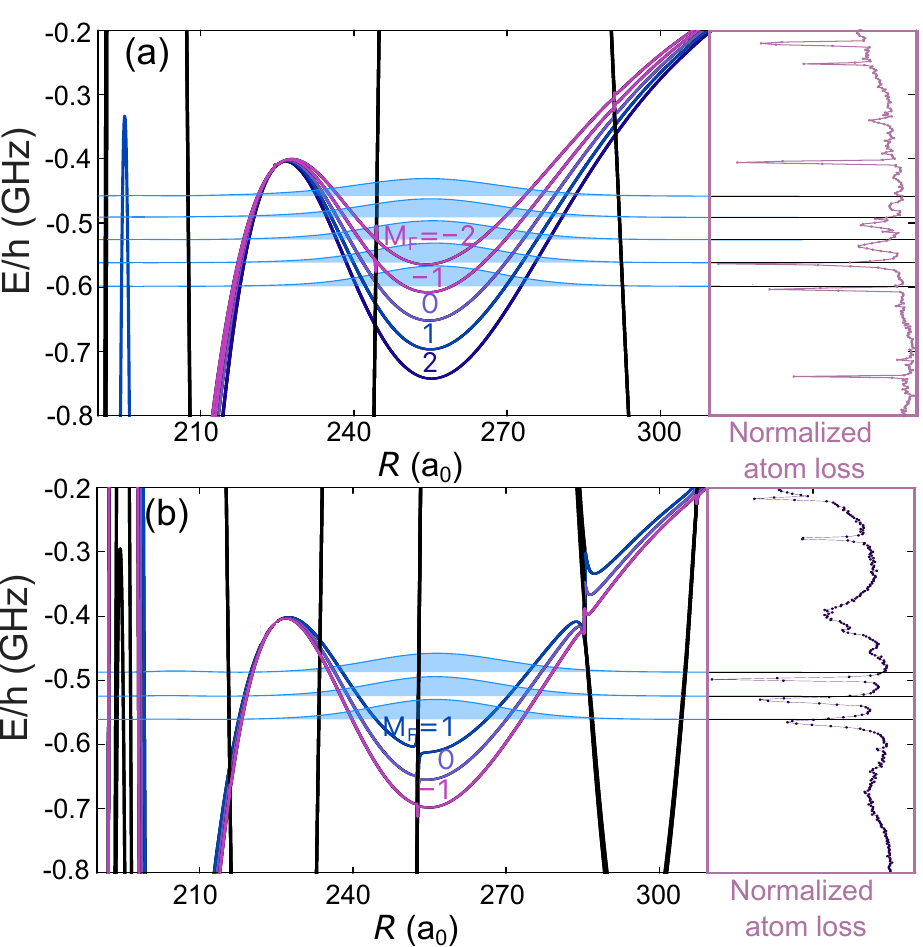}
\caption{(a,b) Calculated $\Pi$-state  PECs (colored lines) correlating with the $15P_{3/2},F=2$ (a) or 
$15P_{3/2},F=1$ (b) atomic pair states. The $\Pi$-state PECs are characterized by $m_j = 3/2$ and the labeled $M_F$ value.  For states with $m_j = -3/2$, the $M_F$ labeling would be inverted.  The vibrational ground state energy and the wavefunction of each PEC are shown in blue. The $\Sigma$ potential curves (black) cut diabatically through both sets of PECs. The experimental spectra are shown to the right of the calculated PECs.
}
\label{fig:PECs}
\end{figure}

Fig. ~\ref{fig:PECs} illustrates the results of such a calculation for the $\Pi$-state of the $15P_{3/2}$ ULRM for $F=2$ (a) and $F=1$ (b). 
The PECs, shown as colored curves and labeled by their dominant $M_F$ character, support
only one vibrational state each. 
The calculated binding energies of these states match those detected in the experimental spectra shown on the right hand side of the PECs.

Furthermore, the calculations in Fig. \ref{fig:PECs} show that the $\Pi$-symmetry PECs are intersected nearly diabatically by the $\Sigma$-symmetry PECs, highlighting the weak coupling between $\Pi$- and $\Sigma$-states.
Our vibrational state calculations assume strictly diabatic $\Pi$-symmetry PECs. 
An in-depth study of the decay processes of the $\Pi$-states, taking into account the effect of non-adiabatic couplings, remains an interesting task for future work  \cite{durst_nonadiabatic_2025,srikumar_vibrationally_2025}. 
The experimental data already reveal tell-tale variations in the signal strengths and linewidths (see, e.g. the spectra in Fig. \ref{fig:PECs}).

An explanation for the observed multiplet structure emerges from a perturbative treatment of the interaction $\hat V_e$ (see also \cite{Supplemental}). We consider the first order shift of the pair states 
$\ket{m_jM_F}\equiv\ket{n\left(1\tfrac{1}{2}\right)\tfrac{3}{2}m_j}\otimes \ket{\left(\tfrac{3}{2}\tfrac{1}{2}\right)FM_F}$, where the unperturbed states with different combinations of $m_j$ and $M_F$ satisfying $\Omega = m_j + M_F$ are degenerate { for a given $\Omega$}. 
The states which diagonalize the perturbation matrix $\bra{m_jM_F}\hat V_e\ket{m_j'M_F'}$ will, in general, be mixtures of $m_j$ and $M_F$ states. 
  However, we find that the states with $|m_j| = 3/2$ are, to a very good approximation, decoupled from those with $|m_j| = 1/2$, due to a far off-resonant
  interaction: The 
energy separation of
diagonal elements with $|m_j| = 3/2$ and
$|m_j| = 1/2$ 
is much larger than the off-diagonal 
coupling elements. 
The binding energy of the $|m_j| = 3/2$ state (i.e. the $\Pi$-state) and its coupling strength to the $|m_j| = 1/2$ state
(i.e. dominantly $\Sigma$-state)
are both based on $p$-wave interactions
which are proportional to the local gradient of the Rydberg wavefunction \cite{hamilton2002shape}.
Since the internuclear axis lies on the nodal line of the $\Pi$-state, the gradient component along $\vec R$ vanishes. Thus, the gradient is perpendicular to $\vec R$, pointing in azimuthal direction. It is very small as the $p$-orbital oscillates only once across its large circumference (see Fig. \ref{fig:intro}).
In comparison, for $|m_j| = 1/2$ the wavefunction gradient is typically much larger. 
For the dominant $\Sigma$-component it points in the radial direction where the wavefunction is quickly oscillating.
This leads to comparatively much larger diagonal energy terms for the $|m_j| = 1/2$  $\Sigma$- dominated states.

To first order in perturbation theory we obtain the $\Pi$-state PECs
 \begin{align}
    U_{m_j,M_F}^{n,F}(R) = \frac{U_{n\Pi}(R)}{8}\Bigg[&\left(5 + \Delta\right)+\frac{2}{3}g_F \left(3 - \Delta\right)m_jM_F
\Bigg]  ,\label{eq:xpotential}
\end{align}
where  $\Delta=[a^3_{(^3P_1)}(K)+2a^3_{(^1P_1)}(K)]/a^3_{(^3P_2)}(K)$, 
 $g_F = (-1)^F/2$ is the Landé g-factor for the perturber, and
\begin{align}
  U_{n\Pi}(R)\equiv  6\pi a_{(^3P_2)}^3(K)\left|\sqrt{\frac{3}{4\pi}}\frac{\psi_{\nu\ell=1}(R)}{R}\right|^2
    \label{eq:pistatepotential}
\end{align}
is the PEC for the fully spin-stretched state with $m_j = 3/2, M_F = 2$. $\psi_{\nu l}$ is the radial Rydberg electron wave function.
Eq.~\ref{eq:xpotential} shows how the spin-dependence of the interaction (quantified by $\Delta$) leads to a splitting of the PECs for different $M_F$.
This splitting is identical to that created by an effective magnetic field $\vec B_\mathrm{eff}\equiv\tfrac{U_{n\Pi}(R)}{12\mu_B}(3-\Delta)\vec j$ acting on the magnetic moment $\vec{\mu}=-g_F\mu_B\vec F$ of the perturber, {where $\mu_B$ is the Bohr magneton}. 
The End Matter presents a derivation of this effective magnetic field for a simplified case with no $J$-dependence in the scattering volumes. 
In the absence of any spin-dependence, $\Delta=3$ and $\vec B_\mathrm{eff}=0$. 
The number of levels depends on the nuclear spin of the perturber and would, for example, grow to $5$ and $7$ (for $F=2$ and $F=3$, respectively) for $^{85}$Rb.

\begin{figure}[b]
\centering
\includegraphics[width=0.4\textwidth]{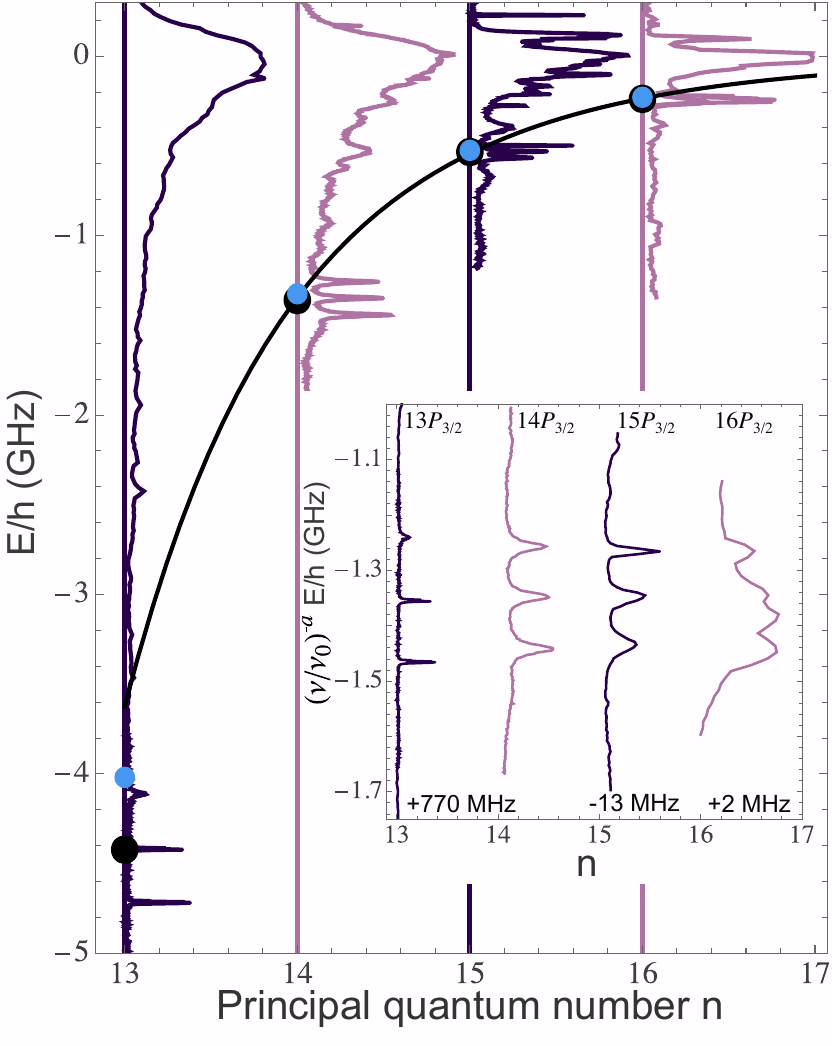}
\caption{
Scaling properties of the $\Pi$-states as a function of $n$.
Shown are measured spectra for  $n = 13$ to $16$ and $F=1$.
A curve $\propto (n-\mu_{P_{3/2}})^a$ is overlaid, where $a=-10.74$ is the fitted exponent. 
The black markers denote the experimentally determined positions of the central peak in the triplet.
The smaller blue markers denote the positions predicted by the full theory.
The inset shows the triplet peaks in closeup, scaled by a factor $(\nu/\nu_0)^{-a}$ where $\nu_0 = 11.355$ is the effective principal quantum number of the $14P_{3/2}$ state.
The spectra for $n=13,15,16$ in the inset are vertically offset by the labeled amounts to align the middle peak for better comparison. } 
\label{fig:scaling}
\end{figure}

Equation \ref{eq:xpotential} reveals the influence of the scattering volumes on the multiplet structure.
This allows for a simple estimate of the accuracy of these parameters.
The comparison between calculated and observed spectra in Figs.~\ref{fig:spectra} and \ref{fig:PECs} shows that 
the positions of the deepest levels (corresponding to $M_F=2$) 
relative to the atomic threshold agree to about $5$\%. 
Using Eq.~\ref{eq:pistatepotential} this  indicates that the employed value of the $^3P_2$ scattering volume is therefore quite accurate over the collisional energy range probed, as this is the only one determining the $M_F=2$ PEC ({as the dependence on $\Delta$ drops out}). 
In contrast, the theoretically predicted splittings are too large by about 10\%.
This reflects the limited accuracy in the determination of the $^3P_1$ resonance position \cite{bahrim2001negative,engelPrecision2019} (and to a lesser extent, the $^1P_1$ scattering volume), as can be seen from the dependence on the splitting in Eq.~\ref{eq:xpotential} on the parameter $\Delta$.

Fig.~\ref{fig:scaling} investigates the scaling properties of the $\Pi$-state ULRM spectra for $13\le n \le 16$ and $F=1$ (for a similar analysis of the $F=2$ spectra see  Fig.~\ref{fig:scaling_2} in the End Matter). 
From a simple theoretical estimate we expect
the  binding energies to decrease as $\nu^{-10}$. This is because the binding energy is proportional to  the azimuthal component of the wave function gradient,
which is proportional to $|\psi_{\nu\ell}(R)/R|^2$. 
At the outermost  anti-node, $|\psi_{\nu\ell}(R)|$ scales as  $\nu^{-3}$ and
the characteristic size of the Rydberg orbital scales as $ \langle R \rangle \sim \nu^2$.
We have fit the energy of the central peak {($M_F = 0$)} in each multiplet (marked with black dots in Fig.~\ref{fig:scaling}) for $14\le n\le16$ to the scaling law $\nu^{a}$, obtaining $a=-10.74$. 
This is {close} to the theoretical estimate which considers only the spatial dependence of the Rydberg wave function. 
The deviation between this and the fitted value reflects the additional $n$-dependence found in the energy-dependence of the $p$-wave scattering volumes as well as non-perturbative effects. 

The inset of Fig.~\ref{fig:scaling} demonstrates the scaling of the 
line splittings of the triplet structures after {dividing by the scaling factor
$[(n - \mu_{P_{3/2}})/\nu_0]^a$}, as the rescaled multiplets exhibit the same splitting. 
The {unscaled} splitting of the $16P_{3/2}$ $\Pi$-state molecules is already so small that it is only partially resolved even with our higher resolution experimental method. 
Extrapolation to higher $n$ using the obtained scaling law predicts a sub-MHz binding energy of the vibrational ground state by the $26P_{3/2}$ level. {This
} helps explain the absence of any observations of the $\Pi$-symmetry multiplet structure in earlier explorations of $nP$-state ULRMs \cite{niederprum_observation_2016,manthey_dynamically_2015,niederprum2016rydberg}. 
Finally, we observe that the $13P_{3/2}$ level is slightly more deeply bound than predicted, both by the fitted scaling law and by our full calculation (blue dot on figure). 
This might indicate that the current theoretical model based on contact pseudopotentials begins to break down in its quantitative description of the full electron–Rb interaction at this low $n$. 
Here, the length scale separation between the Rydberg orbit and the range of the perturber-electron interaction becomes less reliable. 
Further work will be necessary to bridge the gap between the theoretical description of ULRMs at low principal quantum number and quantum chemistry calculations of tightly bound molecules near the electronic ground state. 

In conclusion, facilitated by spectroscopy of low-$n$ Rydberg levels, we have reported the first observation of pure $\Pi$-symmetry ultralong-range Rydberg molecules. 
We have confirmed the $\Pi$-character of these states by detecting their characteristic multiplet structure and by comparing their binding energies across several principal quantum numbers levels to extract their unique scaling with $n$. 
These states can provide a unique advantage in the effort to extract information about scattering phase shifts from ULRM spectroscopy. 
Previous studies \cite{exner_high_2025, desalvo_ultra-long-range_2015, MacLennan2019, Boettcher2016, engelPrecision2019,peper_heteronuclear_2021,sasmannshausen_experimental_2015,legrand2025revealing}  have involved states of predominantly $\Sigma$ character, which always exhibit some coupling to the $s$-wave scattering phase shifts. 
There is no such coupling for pure $\Pi$-symmetry states, and therefore extended spectroscopy of these should provide the sharpest information about the $p$-wave scattering phase shifts.
Because of the low $n$ studied here, the vibrational states probe a broad range of electronic kinetic energies $K$ from $\sim 5$ meV up to $\sim 90$ meV. 
This provides further motivation to continue exploration of ULRMs in the limit of low principal quantum number.

\textit{Acknowledgments--} 
We thank Aditya Dev for assistance developing the computer code used in the Green's function calculations. The Purdue portion of this work was
supported in part by NSF Grant No. PHY-2512984.
JHD would like to acknowledge funding by the German Research Foundation (DFG)
within the priority program “Giant Interactions in Rydberg
Systems” [DFG SPP 1929 GiRyd (Project No. HE 6195/3-1)] and by Q-DYNAMO (EU HORIZON-MSCA-2022- SE-01)
within Project No. 101131418. 


%

\newpage
\section{End Matter }

\subsection{Experimental schemes}
Using the following schemes, we can optimize our experiment for either spectral resolution or sensitivity.

\textit{Scheme I:} The atoms are illuminated by a single pulse of the photoassociation laser 
of well-defined frequency
with a length of typically a few 100 ms, followed by absorption imaging of the remaining number of atoms. This allows for comparatively high frequency resolution, however, the sensitivity for detecting weak signals is limited.

\textit{Scheme II:} For measurements requiring higher detection sensitivity, we employ a modified experimental scheme. In our setup, the dipole trap is spatially overlapped with a linear Paul trap. After photoassociation, a fraction of the ULRMs ionize. These ions are captured in the Paul trap and subsequently collide with the ultracold atomic cloud, thereby opening an additional atom-loss channel, i.e. ion mediated loss amplification takes place. The Paul trap is operated with a radio-frequency (RF) electric field oscillating at $4.2~\mathrm{MHz}$.
The electric field induces time-dependent Stark shifts of the atomic Rydberg states and causes asymmetric spectral broadening toward {smaller} frequencies. To mitigate this effect, the UV excitation is carried out in a sequence of 25 ns-long pulses which occur whenever the RF field of the Paul trap vanishes.
This substantially suppresses Stark-induced broadening \cite{Haze_broadening_2019, Ewald2019} 
and reduces atom-number loss associated with photon scattering on the Rydberg-atom transition.
However, this is at the expense of an additional pulse-induced spectral broadening of up to $50~\mathrm{MHz}$.

We used Scheme II for the measurements of Fig. \ref{fig:spectra}. Scheme I was used for the measurements of Figs. \ref{fig:PECs}, \ref{fig:scaling}, and for all measurements shown in Fig. \ref{fig:scaling_2} except for the $13P_j$, $F=2$ data.

\subsection{Full experimental spectra}

\begin{figure}[t]
\centering
\includegraphics[width=0.45\textwidth]{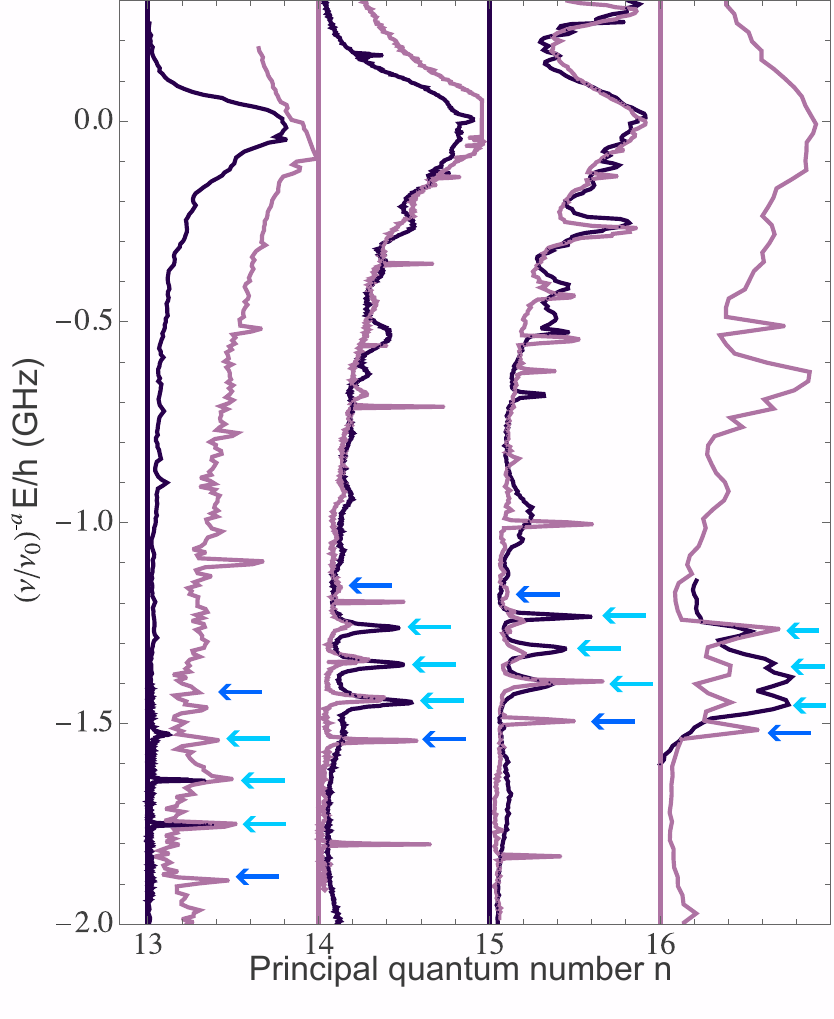}
\caption{ Measured spectra for $\Pi$ states of the $n P_{3/2}$ ULRMs, where $n=13-16$, for $F = 1$ (dark color) and $F=2$ (light color). 
The energies have been scaled by the factor $(\nu/\nu_0)^{-a}$, where $a=-10.74$, $\nu=n - \mu_{P_{3/2}}$, 
and $\nu_0 = 11.355$. 
The $F=1$ data is also shown in Fig. \ref{fig:scaling}. 
The light blue arrows mark the positions of the triplet energies in the $F=1$ spectrum as well as the positions of the $m_F = 0, \pm 1$ lines of the quintet  $F=2$ spectrum, while the dark blue arrows mark the $m_F = \pm 2$ peaks of the quintet $F=2$ spectrum. 
The shallowest quintet level is much broader than the others, and in the $16P_{3/2}$ spectrum it is not measured.}

\label{fig:scaling_2}
\end{figure}

Figure \ref{fig:scaling_2} shows the observed spectra for $nP_{3/2}$ Rydberg states, where $n$ ranges between  13 and 16. 
 These data were taken without the Paul trap to better resolve the multiplet structure for $15P_{3/2}$ and $16P_{3/2}$.  
The energies are defined as in the inset to Fig.~\ref{fig:scaling} to
show the near degeneracy of the triplet $F=1$ series with the
$m_F = 0,\pm1$ states of the quintet $F=2$ series.

Table \ref{tab:dataset} lists the binding energies of all detected molecular states. The theoretical positions and widths were calculated in the same way as in Refs.~\cite{durst_nonadiabatic_2025,srikumar_vibrationally_2025}.
The theoretical decay width $\Gamma$ provides information only about the stability of the molecule with respect to pre-dissociation driven by tunneling of the vibrational bound state towards smaller internuclear distances. It is thus an underestimate of the decay width as spontaneous decay of the Rydberg state, collisional decay, and other processes are ignored. The Franck-Condon Factor $
    \mathcal{F}=\int \psi_i(R)\psi( R)\dd{R}$
was calculated for the ground vibrational 
ULRM states $\psi(R)$ bound in the PECs, taking the initial pair state $\psi_i(R)$ to be a scattering state of two $^{87}$Rb atoms at a temperature of 1 $\mu$K. Because the scattering wavefunction $\psi_i(R)$ is box-normalized, the
Franck-Condon factors are only defined up to an unknown global factor.

\subsection{Simplified treatment of multiplet structure}
As discussed in the main text and Supplemental Material \cite{Supplemental}, the electronic Hamiltonian is approximately diagonal in the uncoupled representation $|m_jM_F\rangle \equiv |n(\ell s_e)jm_j\rangle\otimes|FM_F\rangle$, where $|m_j|=3/2$ and $-F\le M_F \le F$. The diagonal matrix elements (with the threshold energy of the atomic pair defined as zero and neglecting for simplicity the $J$ dependence of the $^3P$ scattering volumes) are
\begin{align}
    H_{m_j,M_F} &= \sum_{S=0,1}\bra{m_j,M_F}V_S \hat P_{S}\ket{m_j,M_F},
    \label{eq:hmjmf}
\end{align}
    where 
    \begin{align}
   \hat P_{S}&=({2S+1})/{4} +(-1)^{S+1} \vec s_e\cdot \vec s_p
   \end{align}
   is a projection operator onto a state of total spin $S$ and 
   \begin{align}
   V_S&\equiv    6\pi a^3_{(^{2S+1}P)}(K)\cdot\frac{3}{4\pi}\left|\frac{\psi_{\nu\ell=1 }(R)}{R}\right|^2
\end{align}
is the spin-dependent interaction potential with $\psi_{\nu \ell}(R)$  the radial Rydberg wave function.
We rewrite 
Eq. \ref{eq:hmjmf} as
\begin{align}
   H_{m_j,M_F}={(3V_1 + V_0)}/{4}+ \bra{m_jM_F}\Delta_V\vec s_e \cdot \vec s_p\ket{m_jM_F}\nonumber,
\end{align}
i.e. a spin-averaged electron-perturber interaction plus a spin-spin coupling term proportional to the difference between spin-dependent interactions $\Delta_V = (V_1 - V_0)$.

Using the projection theorem \cite{sakurai2020modern}, we have (in this $2(2F+1)$-dimensional magnetic subspace) the relations
\begin{align*}
    \vec s_e  &=\frac{\langle \vec s_e\cdot\vec j\rangle} {j(j+1)} \vec j= \frac{j(j+1) + s_e(s_e + 1) - \ell(\ell + 1)}{2j(j+1)}\vec j;\\
    \vec s_p &=\frac{\langle \vec s_p\cdot\vec F\rangle} {F(F+1)} \vec F = \frac{F(F+1) + s_p(s_p + 1) - i(i+1)}{2F(F+1)}\vec F.
\end{align*}
For the $nP_{3/2}$ Rydberg state and $^{87}$Rb perturber with nuclear spin $i=3/2$, we have $\vec s_e=\tfrac{1}{3}\vec j$ and $\vec s_p = \tfrac{(-1)^F}{4}\vec F$ and thus 
\begin{align}
    (V_1 - V_0)\vec s_e \cdot \vec s_p &= (V_1 - V_0)\frac{(-1)^F}{12}\vec j\cdot\vec F.
\end{align}
Thus, $\bra{jm_j,FM_F}\vec j\cdot \vec F \ket{jm_j,FM_F}= m_j M_F$ and the resulting PECs obtained in first-order perturbation theory are
\begin{align}
\label{eq:u}
   U_{m_j,M_F}(R)&= \frac{1}{4}\left((3V_1 + V_0)+\frac{\Delta_V}{3}(-1)^Fm_jM_F\right)\\
   &= \frac{V_1}{8}\left(6 + \frac{2V_0}{V_1}+\frac{4}{3}\left(1-\frac{V_0}{V_1}\right)g_Fm_jM_F\right),\nonumber
\end{align}
where $g_F = (-1)^F/2$. 
Eq. \ref{eq:u} is equivalent to Eq.~\ref{eq:xpotential} after setting
 $\Delta=1+2a^3_{(^1P_1)}(K)/a^3_{(^3P_2)}(K)$.

\begin{table*}[hb]
\centering
\caption{Measured (exp) and calculated (th) $nP_{3/2}$ molecular {$\Pi$-}state energies ($n=13$--$16$) referenced to the corresponding atomic asymptotes. The
 binding energy $\Delta E$ is in GHz$\times h$. The typical experimental uncertainty is approximately 2~MHz; n.o.: not observed. The calculated widths $\Gamma$ are in MHz$~\times h$ 
and the Franck-Condon Factor $\mathcal{F}$ is multiplied by $1000$ for readability.
}
\setlength{\tabcolsep}{8pt}
\renewcommand{\arraystretch}{1.15}
\begin{tabular}{c c | c}
\toprule
$n$ & $F=1$ & $F=2$ \\
\midrule
13 &
\begin{tabular}[t]{@{}c c c c c c@{}}
$M_F$ & $\Omega$  & $\Delta E_\mathrm{exp}$ & $\Delta E_\mathrm{th}$ & $\Gamma$ & $1000\times|\mathcal{F}|$\\ \midrule
\\
1 & 5/2 & -4.718 & -4.335 & $<10^{-4}$ & 0.027 \\
0 & 3/2 & -4.422 & -4.019  & 0.009 & 0.037\\
-1 & 1/2 & -4.110 & -3.675 & 0.275 & 0.037\\
\end{tabular}
&
\begin{tabular}[t]{@{}c c c c c c@{}}
$M_F$ & $\Omega$ & $\Delta E_\mathrm{exp}$ & $\Delta E_\mathrm{th}$ & $\Gamma$ & $1000\times|\mathcal{F}|$\\  \midrule
-2 & -1/2 & -4.984 & -4.659 & 0.289 & 0.037\\
-1 & 1/2 & -4.609 & -4.228 & 0.865 & 0.038\\
0 & 3/2 & -4.294 & -3.871 & 1.134 & 0.039\\
1 & 5/2 & -4.034 & -3.550 & 4.232 & 0.039\\
2 & 7/2 & -3.828 & -3.245 & 6.420 & 0.040\\
\end{tabular}
\\
\midrule
14 &
\begin{tabular}[t]{@{}c c c c c  c@{}}
$M_F$ & $\Omega$ & $\Delta E_\mathrm{exp}$ & $\Delta E_\mathrm{th}$ & $\Gamma$ & $1000\times|\mathcal{F}|$\\  \midrule
\\
1 & 5/2 & -1.453 & -1.424 & 0.004 &0.113\\
0 & 3/2 & -1.359 & -1.324 & 1.748 & 0.115\\
-1 & 1/2 & -1.268 & -1.226 & 2.201 & 0.116\\
\end{tabular}
&
\begin{tabular}[t]{@{}c c c c c  c@{}}
$M_F$ & $\Omega$ & $\Delta E_\mathrm{exp}$ & $\Delta E_\mathrm{th}$ & $\Gamma$ & $1000\times|\mathcal{F}|$\\  \midrule
-2 & -1/2 & -1.556 & -1.525 & 2.391 & 0.117\\
-1 & 1/2 & -1.451 & -1.413 & 2.567 & 0.117\\
0 & 3/2 & -1.353 & -1.307 & 4.461 & 0.119\\
1 & 5/2 & -1.266 & -1.207 & 5.485 & 0.123\\
2 & 7/2 & -1.171 & -1.105 & 12.64 & 0.128\\
\end{tabular}
\\
\midrule
15 &
\begin{tabular}[t]{@{}c c c c c c@{}}
$M_F$ & $\Omega$ & $\Delta E_\mathrm{exp}$ & $\Delta E_\mathrm{th}$ & $\Gamma$ & $1000\times|\mathcal{F}|$\\  \midrule
\\
1 & 5/2 & -0.570 & -0.561 & 0.260 & 0.215\\
0 & 3/2 & -0.534 & -0.525 & 0.307 & 0.225\\
-1 & 1/2 & -0.502 & -0.487 & 0.444 & 0.230\\
\end{tabular}
&
\begin{tabular}[t]{@{}c c c c c c@{}}
$M_F$ & $\Omega$ & $\Delta E_\mathrm{exp}$ & $\Delta E_\mathrm{th}$ & $\Gamma$ & $1000\times|\mathcal{F}|$\\  \midrule
-2 & -1/2 & -0.602 & -0.598 & 0.485 & 0.242\\
-1 & 1/2 & -0.562 & -0.562 & 1.92 & 0.249\\
0 & 3/2 & -0.535 & -0.526 & 4.16 & 0.255\\
1 & 5/2 & -0.501 & -0.491 & 11.0 & 0.257\\
2 & 7/2 & -0.472 & -0.458 & 11.1 & 0.285\\
\end{tabular}
\\
\midrule
16 &
\begin{tabular}[t]{@{}c c c c c c@{}}
$M_F$ & $\Omega$ & $\Delta E_\mathrm{exp}$ & $\Delta E_\mathrm{th}$ & $\Gamma$ & $1000\times|\mathcal{F}|$\\  \midrule
\\
1 & 5/2 & -0.250 & -0.234 & 1.81 & 0.264\\
0 & 3/2 & -0.238 & -0.229 & 4.12 & 0.265 \\
-1 & 1/2 & -0.218 & -0.214 & 5.77 & 0.307\\
\end{tabular}
&
\begin{tabular}[t]{@{}c c c c c c@{}}
$M_F$ & $\Omega$ & $\Delta E_\mathrm{exp}$ & $\Delta E_\mathrm{th}$ & $\Gamma$ & $1000\times|\mathcal{F}|$\\ \midrule
-2 & -1/2 & -0.264&  -0.258& 7.23  & 0.309\\
-1 & 1/2 & -0.253&  -0.248& 9.92  & 0.394\\
0 & 3/2 & -0.235&  -0.221& 10.00 & 0.398\\
1 & 5/2 & -0.220&  -0.206& 12.24 & 0.429\\
2 & 7/2 &   n.o.&  -0.187& 12.58 & 0.522\\
\end{tabular}
\\
\bottomrule
\end{tabular}
\label{tab:dataset}
\end{table*}

\clearpage

\title{Supplementary information: Observation of $\Pi$-symmetry ultralong-range Rydberg molecules}
\author{Author Name}
\affiliation{Department, University}
\date{\today}
\maketitle
\onecolumngrid
\renewcommand{\thefigure}{S\arabic{figure}}
\renewcommand{\thetable}{S\arabic{table}}
\renewcommand{\thesection}{S\arabic{section}}
\setcounter{figure}{0}
\setcounter{section}{0}
\setcounter{table}{0}

\section{High-resolution $14P_{3/2}$ spectrum}
\begin{figure*}[hb]
\centering
\includegraphics[width=\textwidth]{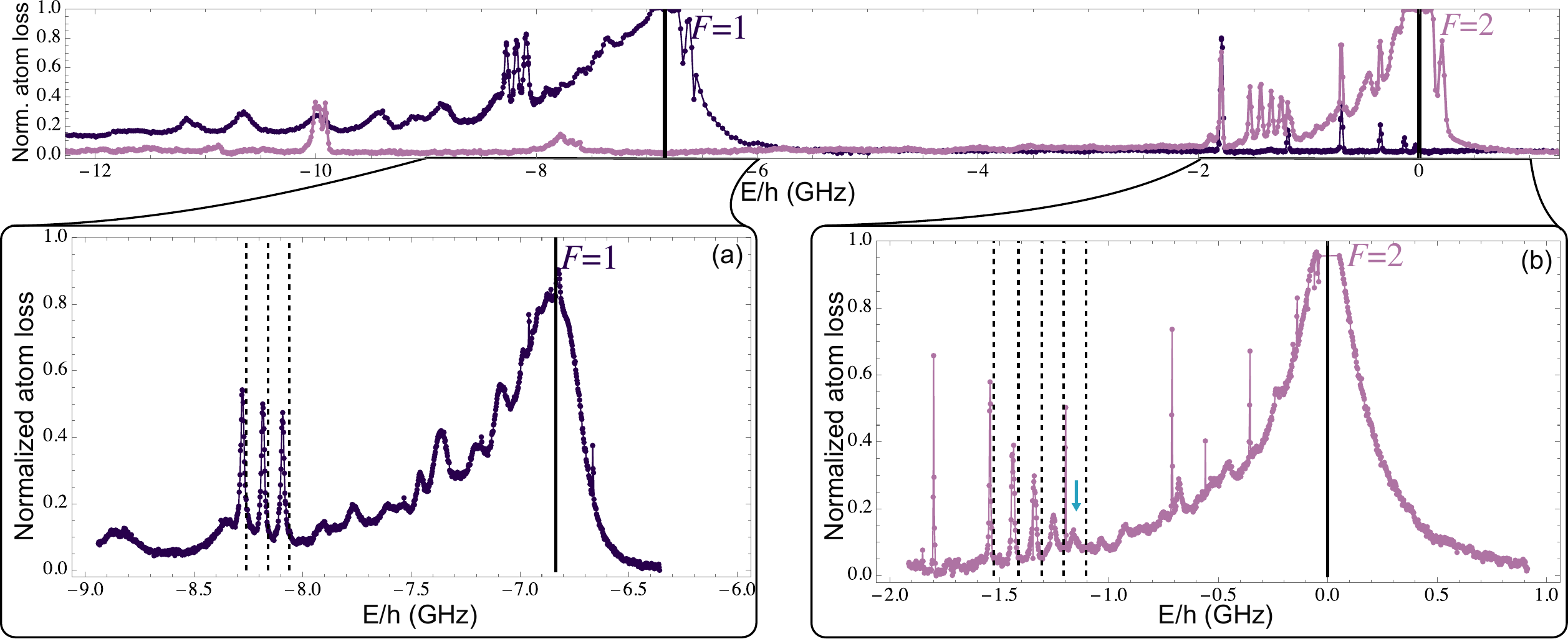}
\caption{ULRM spectra in the vicinity of the $14P_{3/2}$ Rydberg state for $^{87}$Rb atoms initially prepared in the $F=1$ (dark, purple) and $F=2$ (light, pink) state. Shown is the normalized atomic loss as a function of the detuning frequency of the UV spectroscopy laser from the $5S_{1/2}(F=2) \rightarrow 14P_{3/2}$ atomic transition. The solid vertical lines denote the $F=1$ and $F=2$ atomic thresholds.
In order to easily identify Rydberg levels  which exhibit 
mixtures of $F = 1$ and 2, and which are therefore not
$\Pi$-states,
the spectra for atoms initially prepared in the $F=1$ ground state were shifted by twice the ground-state hyperfine splitting $(-6.834\, \mathrm{ GHz} \times \mathrm{h})$.
Then the  $F = 1$  and $F = 2$ signals of these lines coincide. 
The insets {(a) and (b)} show zooms into the spectral regions around the spectra of the $\Pi$ molecular states. In these panels, the spectra were taken with the Paul trap turned off. The positions, splittings, and number of states observed are in convincing agreement with theoretical calculations 
(dashed vertical lines). 
The blue arrow marks the fifth level of the $F=2$ series.
}
\label{fig:spectra_highres}
\end{figure*}

For completeness, Fig. ~\ref{fig:spectra_highres} is a copy of Fig. 2 from the main text, except that the insets (a) and (b)
show the $14P_{3/2}$ molecular spectra taken without using the Paul trap. 
The resulting spectrum's high resolution confirms that the fifth level in the $F=2$ multiplet is indeed the shoulder of the peak seen in Fig.~2; here it is {observed} as a separate peak blue-detuned from the very narrow peak at about -1.2 GHz. This {narrow peak}, as seen in Fig. 2 in the main text, can be found in both 
{the $F = 1$ and $F = 2$
spectra} and therefore has substantial mixing of both $F$ components. Therefore, it cannot be a $\Pi$-symmetry state where $F$ is approximately a good quantum number. Fig.~\ref{fig:spectra_highres} also shows that the {lineshape of the narrow peak} is very different from the other levels in the multiplet, indicating its different origin: a vibrational level bound in one of the much deeper wells of the $\Sigma$-dominated potential energy curves. 

\section{Details of the perturbative treatment}
As discussed in the main text, we use a perturbative treatment of the Fermi-Omont pseudopotential to obtain a basic understanding of the molecular potential energy curves. While this perturbative treatment is not sufficiently accurate to yield molecular binding energies that we can compare directly with the peak positions measured experimentally, it gives a qualitative picture of the mechanism by which $\Pi$- and $\Sigma$-dominated states separate energetically and of the observed multiplet structure. 

The unperturbed pair-state wave function, $\Psi_{n(\ell s_e)jm_j}(\vec{r})\ket{FM_F}=\langle\vec r\ket{n(\ell s_e)jm_j}\otimes \ket{FM_F}$, is conveniently written in terms of a partial wave expansion around the perturber, given by (following Eq. (16) of  \cite{eilesHamiltonian2017})
\begin{equation}
    \Psi_{\nu(\ell s_e)jm_j}(\vec{r})\ket{FM_F}=\sum_{m_e=-s_e}^{{m_e=}s_e}\sum_{L=0}^1\sum_{M_L=-L}^L X^L f_L C^{jm_j}_{\ell M_L,s_em_e}Q_{LM_L}^{n\ell j}(R)Y_{LM_L}(\hat{X})\chi_{m_e}^{s_e}\ket{FM_F},
\end{equation}
where $\ell$ and $L$ are the electronic orbital angular momenta with respect to the Rydberg core and the perturber, respectively, $C_{\ell M_L,s_em_e}^{jm_j}$ denotes a Clebsch Gordan coefficient, $\vec X = \vec r - \vec R$ denotes the electron's position relative to the perturber, and additionally
\begin{align}
Q_{L=0,M_L=0}^{n\ell j}(R)&=\sqrt{\frac{2\ell + 1}{4\pi}}\psi_{\nu\ell}(R)\\
Q_{L=1,M_L=0}^{n\ell j}(R)&=\sqrt{\frac{2\ell + 1}{4\pi}}\frac{d}{d R}\psi_{\nu\ell}(R)\\
    Q_{L=1,M_L=\pm1}^{n\ell j}(R)&=\sqrt{\frac{(2\ell + 1)(\ell +1)\ell }{8\pi}}\frac{\psi_{\nu\ell} (R)}{R}. 
\end{align}
In these expressions, $\psi_{\nu\ell(R)}=\frac{u_{\nu\ell}(R)}{R}$ is the radial Rydberg wave function for the state with effective principal quantum number $\nu = n - \mu_{l_j}$.  
The Fermi-Omont pseudopotential valid for spin-dependent scattering phase shifts can be written, including the identity operator in the nuclear spin degree of freedom $\sum_{m_i}\ket{im_i}\bra{im_i}$ (following Eq. (13) of  \cite{eilesHamiltonian2017}), as  
\begin{align*}
    \hat{V}_e(\vec r,\vec R)&=\sum_{(L,S)JM_J,m_i}\ket{(LS)JM_J,im_i}\frac{(2L+1)^2}{2}[a_{(^{2S+1}L_J)}(K)]^{2L+1}\frac{\delta(X)}{X^{2(L+1)}}\bra{(LS)JM_J,im_i}\\&=\sum_{\{\alpha\}}\ket{\alpha}\frac{(2L+1)^2}{2}[a_{(^{2S+1}L_J)}(K)]^{2L+1}\frac{\delta(X)}{X^{2(L+1)}}\bra{\alpha},
\end{align*}
using the shorthand 
$\ket{\alpha}=\ket{(L,S)JM_J,im_i}$ where $J$ is the total angular momentum of the Rydberg electron together with the ground-state atom's valence electron, $M_J$ its projection onto the $z$ axis, and $S$ the total electron spin. 
$\delta(X)$ is the Dirac delta function. 
It is then straightforward to obtain the matrix elements as a sum over Clebsch-Gordan coefficients
\begin{align*}
   &\bra{n(\ell s_e)jm_j',FM_F'}\hat{V}_e\ket{n (\ell s_e)jm_j'',FM_F''}\\&\,\,\,\,\,\,\,\,\,\,\,\,\,\,\,\,\,\,\,\,=2\pi\sum_{\{M'\}}\sum_{\{M''\}}\sum_{\{\alpha\}} C^{jm_j'}_{\ell M_L',s_em_e'}C^{SM_S'}_{s_em_e',s_pm_p'}C^{JM_J}_{LM_L',SM_S'}C^{FM_F'}_{s_pm_p',im_i}Q_{LM_L'}^{n\ell j}(R)\\&\,\,\,\,\,\,\,\,\,\,\,\,\,\,\,\,\,\,\,\,\times(2L+1)[a_{(^{2S+1}L_J)}(K)]^{2L+1}C^{jm_j''}_{\ell M_L'',s_em_e''}C^{SM_S''}_{s_em_e'',s_pm_p''}C^{JM_J}_{LM_L'',SM_S''}C^{FM_F''}_{s_pm_p'',im_i}Q_{LM_L''}^{n\ell j}(R),
\end{align*}
where $\{M\}$ is used to denote the collection of magnetic quantum numbers $\{m_e,m_p,M_L,M_S\}$.
Some explicit examples follow. 
For the pair state with $j = 3/2$, $F = 2$, and $\Omega = 7/2$, we obtain the single matrix element
\begin{equation}
   U_{F=2,\Omega=7/2}(R)\equiv \bra{n(\ell s_e)jm_j,FM_F}\hat{V}_e\ket{n (\ell s_e)jm_j,FM_F}= 6\pi a_{(^3P_2)}^3(k) \left(Q_{11}^{nlj}(R)\right)^2.
   \label{eq:pistate}
\end{equation}
Note that this is the same as Eq.~(3) in the main text. 
$\Omega=7/2$ implies that this is the spin-stretched configuration with $M_F = 2$ and $m_j = 3/2$. 
Clearly this spin-stretched state has $\Lambda = 1$. Our question in the following is if pure $\Pi$-symmetry states are also present even when $\Omega$ is not its maximal value. 
The example of $\Omega=5/2$ suffices to illustrate the general structure found for all other $\Omega$ levels. In the following, we continue to consider the interaction as a perturbation to the pair state with $j=3/2$ and $F=2$. 
We obtain, using the short-hand notation $\ket{m_j,M_F}\equiv |n(\ell s_e)jm_j\rangle\otimes\ket{FM_F}$, the matrix:
    \begin{align}
\underline{U}_{2,5/2}(R)&=
\begin{pmatrix}
    \bra{\frac{3}{2},1}\hat{V}_p\ket{\frac{3}{2},1} & \bra{\frac{1}{2},2}\hat{V}_p\ket{\frac{3}{2},1}\\
    \bra{\frac{3}{2},1}\hat{V}_p\ket{\frac{1}{2},2} & \bra{\frac{1}{2},2}\hat{V}_p\ket{\frac{1}{2},2}
\end{pmatrix}\nonumber\\
&=6\pi \begin{pmatrix}
P\left(\frac{1}{8}a_{(^1P_1)}^3(K) + \frac{1}{16}a_{(^3P_1)}^3(K) + \frac{13}{16}a_{(^3P_2)}^3(K)\right)
& 
P\left(-\frac{a_{(^1P_1)}^3(K)}{4\sqrt{3}} + \frac{a_{(^3P_1)}^3(K)}{8\sqrt{3}} + \frac{a_{(^3P_2)}^3(K)}{8\sqrt{3}} \right)
\\
P\left(-\frac{a_{(^1P_1)}^3(K)}{4\sqrt{3}} + \frac{a_{(^3P_1)}^3(K)}{8\sqrt{3}} + \frac{a_{(^3P_2)}^3(K)}{8\sqrt{3}} \right)
& 
V_{22}
\end{pmatrix}
\nonumber
\end{align}
with $P = \left(Q_{11}^{n\ell j}(R)\right)^2$ and
\begin{align*}
    V_{22}=\Bigg[&\frac{2}{9}a_{(^3S_1)}^3(K)\left(Q_{00}^{n\ell j}(R)\right)^2 + \frac{1}{3}a_{(^3P_1)}^3(K)\left(Q_{10}^{n\ell j}(R)\right)^2 + \frac{1}{3}a_{(^3P_2)}^3(K)\left(Q_{10}^{n\ell j}(R)\right)^2  \\&+\frac{1}{6}a_{(^1P_1)}^3(K)\left(Q_{11}^{n\ell j}(R)\right)^2 + \frac{1}{12}a_{(^3P_1)}^3(K)\left(Q_{11}^{n\ell j}(R)\right)^2 + \frac{1}{12}a_{(^3P_2)}^3(K)\left(Q_{11}^{n\ell j}(R)\right)^2\Bigg].
\end{align*}
Note that the presence of an off-diagonal coupling term means that $m_j$ is not, strictly speaking, a good quantum number and hence the perturbed eigenstates (obtained upon diagonalizing this matrix) will be superpositions of electronic states with  $\Lambda=0$ as well as $\Lambda = 1$. 

\begin{figure*}[b]
\centering
\includegraphics[width=0.8\textwidth]{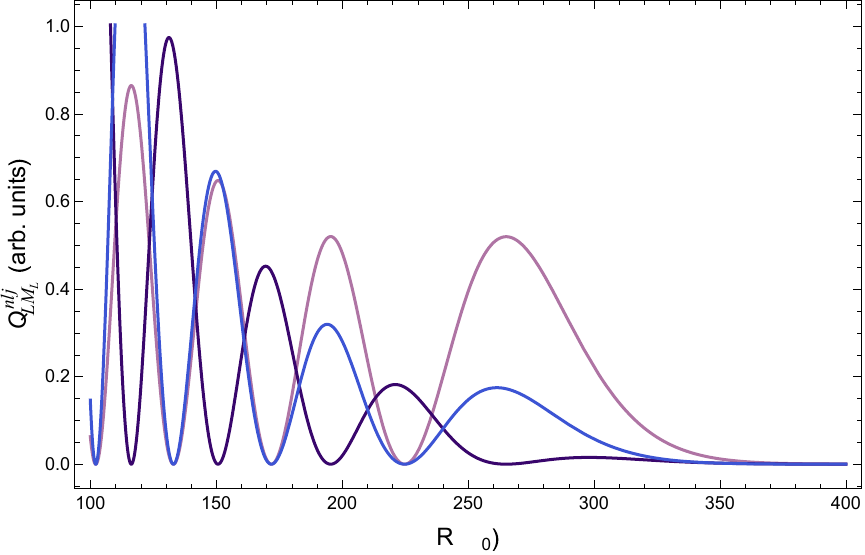}
\caption{ Dependence of the $Q_{LM_L}^{n \ell j}(R)$ terms as a function of $R$. These curves have been scaled by $\nu^6$ (for $L=M_L=0$, shown in pink), $\nu^8$ (for $L=1$, $M_L=0$, shown in purple), and $\nu^{10}$ (for $L=|M_L|=1$, shown in blue). Away from nodes where they vanish, each of the individual terms is, to a good approximation, a factor of $\nu^2$ larger than the next. The curves shown are for $\ell=1$, $j=3/2$, and $\nu=12.355$.   }
\label{fig:Qscaling}
\end{figure*}

However, we notice that the $V_{11}$ and $V_{12}$ matrix elements depend only on $Q_{11}^{n\ell j}(R)$ while $V_{22}$ depends also on $Q_{00}^{n\ell j}(R)$ and $Q_{10}^{n\ell j}(R)$. These latter terms are 2-3 orders of magnitude larger at large $R$ than $Q_{11}^{n\ell j}(R)$, see Fig.~\ref{fig:Qscaling}.  We note also that the off-diagonal elements are proportional to the difference between the spin-orbit averaged triplet and singlet scattering volumes. For low electronic energies $K$ where the electron does not easily penetrate the angular momentum barrier and exchange terms are weak, these couplings are small.   It is therefore generally true that $V_{22}\gg V_{11}\gg V_{12}$. 

Under these conditions, the eigenvalues of $\underline{U}_{2,5/2}(R)$ are approximately 
\begin{align}
  U_{1}&=  V_{11}-\frac{V_{12}^2}{V_{22}}\\
  U_2 &= V_{22}+\frac{V_{12}^2}{V_{22}},
\end{align}
from which we see that the diagonals of the matrix approximate the eigenvalues to an accuracy of approximately one part in $\nu^2$, or better than one percent over the range of $\nu$ considered here.  The PEC $V_{22}(R)$ gives the deep potential energy curve which is predominantly $\Sigma$ symmetry, while the upper diagonal has purely $\Pi$ contributions to first order.

Although we do not show here the full matrix elements for $|\Omega|=3/2$ and $1/2$, they share the same structure. In all cases, the off-diagonal matrix elements (for example, $\bra{m_j=1/2,M_F=0}\hat V_p\ket{m_j=3/2,M_F=-1}$) are proportional to $P$ and vanish if the scattering phase shifts are independent of the total electron spin.
The diagonal elements, on the other hand, are proportional to $P$ if $|\Lambda| = 1$ (for example, for $\bra{m_j = 3/2,M_F=-1}\hat V_p\ket{m_j = 3/2,M_F=-1}$) and are proportional to a combination of $Q_{00}^{n \ell j}(R)$ and $Q_{10}^{n \ell j}(R)$ terms, which are larger than $P$ by a factor $n^2$ or larger, when $\Lambda = 0$ (for example, for $\bra{m_j = 1/2,M_F=0}\hat V_p\ket{m_j = 1/2,M_F=0}$). 
Hence, we can focus only on the diagonal elements and neglect the coupling when focusing on the shallow $|m_j|=3/2$ potential energy curves.

\end{document}